\documentclass{article}
\usepackage{epsfig}
\usepackage{amsmath}
\usepackage{amsthm}
\usepackage{amssymb}
\usepackage{amsfonts}
\usepackage{amstext}
\newtheorem{thm}{Theorem}

\newtheorem{rem}[thm]{Remark}
\newtheorem{prop}[thm]{Proposition}

\newcommand{\be}{\begin{equation}}
\newcommand{\ee}{\end{equation}}
\newcommand{\ba}{\begin{eqnarray}}
\newcommand{\ea}{\end{eqnarray}}
\newcommand{\nit}\noindent
\newcommand{\D}\partial
\newcommand{\nn}{\nonumber}
\newcommand{\cN}{{\cal N}}

\newcommand{\cH}{{\cal H}}
\newcommand{\cA}{{\cal A}}
\newcommand{\g}{{\gamma}}

\def\R{ {\rm R \kern -.31cm I \kern .15cm}}
\def\Rsmall{ {\rm R \kern -.25cm I \kern .11cm}}
\def\C{ {\rm C \kern -.15cm \vrule width.5pt \kern .12cm}}

\numberwithin{equation}{section}

\begin{document}
\title{Geometric Analysis of Bifurcation and  Symmetry Breaking in a 
  Gross-Pitaevskii 
equation }
\date{\today}
\author{R.K. Jackson
\thanks{supported by a grant from the National Science Foundation under award number DMS-0202542} 
\thanks{Department of Mathematics and Statistics, Boston University} 
\hspace{.05in} and M.I. Weinstein 
\thanks{Mathematical Sciences Research, Bell Laboratories - Lucent Technologies, Murray Hill, NJ}
}

\maketitle
\centerline{\it Dedicated to Elliott H. Lieb on the occasion of his seventieth birthday}
\bigskip

\begin{abstract}
 Gross-Pitaevskii and nonlinear Hartree equations are equations of nonlinear Schr\"odinger
 type, which play an important role in the theory of Bose-Einstein condensation, 
 Recent results of Aschenbacher {\it et al} \cite{AFGST} demonstrate, for a class of  
 $3-$ dimensional models, 
   that for large
boson number (squared $L^2$ norm),\  $\cN$,
  the ground state does not have the symmetry properties as  
 the ground state at small $\cN$. 
We present a detailed global study of the symmetry breaking bifurcation for a $1-$ dimensional
 model Gross-Pitaevskii equation, in which the external potential (boson trap) is an attractive
double-well, consisting of two attractive Dirac delta functions concentrated at distinct points.
  Using dynamical systems methods, we present a geometric analysis of the symmetry breaking
 bifurcation of an asymmetric ground state and the exchange of dynamical stability 
 from the symmetric branch
 to the asymmetric branch at the bifurcation point. 
\end{abstract}
\bigskip

\section{Introduction}
The experimental realization of Bose-Einstein condensation has given great
impetus to the study of the Gross-Pitaevskii and nonlinear Hartree equations.
These are equations of {\it nonlinear Schr\"odinger (NLS)} type, 
 with a potential 
having both linear and nonlinear parts.  Such equations arise in the study of large dilute 
systems of bosons; $N$-body quantum mechanics of coupled bosons, where $N$ tends to infinity 
and the coupling strength tends to zero, \cite{DSPS,LSY,Erdos-Yau,Bardos-etal}. 
Of fundamental interest are the properties of the ground state of such systems.
Because of the nonlinearity, the ground state can exhibit non-trivial transitions 
in structure.

Indeed, that such a transition takes place was established in the recent
 paper \cite{AFGST}. 
They consider the nonlinear Hartree energy functional
\be
H_g[\Psi,\bar\Psi] = 
 \frac{1}{2} \int_{\Rsmall^d}|\nabla \Psi(x)|^2 + v(x)|\Psi(x)|^2
+ g |\Psi(x)|^2 \left(V*|\Psi|^2\right)(x)\ \mathrm{d} x.
\label{eq:nlHartree}
\ee
Here, $v(x)$ is an external potential (the boson trap), and $V(x)$ is the two-body
Coulomb interaction potential $V(x)=|x|^{-1}$ between bosons. 
  The ground state of a system of $\cN$ bosons is characterized by 
 solution to the minimization problem
\be
\inf\{H_g[\Psi,\bar\Psi]: \|\Psi\|_2^2=\cN\}.
\label{eq:nlHartree-gs}
\ee
For spatial dimensions $d\ge2$ and the case in which $g<0$ is fixed, 
corresponding to attractive interatomic forces (negative scattering length), 
it is shown that for a class of potentials $v(x)$, {\it e.g.}, the double well 
potential, that if  $\cN$ is sufficiently large, then symmetry breaking in the 
ground state occurs, in the sense that the ground state of (\ref{eq:nlHartree-gs}) 
does not have the same symmetries as $v(x)$.  The proof in \cite{AFGST} is based upon 
the intuition in the double-well case, for example, that for small amplitude states 
($\cN$ small) the nonlinear potential is negligible, and therefore the ground state 
should be a small distortion of the linear ground state ($g=0$), and have the same 
symmetries as the linear ground state.  However, by careful construction of trial functions, 
one can see that for $\cN$ sufficiently large, $\cN>\cN_{\rm thresh}$, it is energetically 
preferable for the ground state to be concentrated in one well or the other,
  but not equally in both. 

The size of $\cN_{\rm thresh}$ and the {\it dynamical} stability properties
 of the various coexisting states 
(symmetric and asymmetric) is unaddressed. In this work we address these questions in the context 
of the simpler model, the one-dimensional Gross-Pitaevskii equation (GP/NLS)
\be
 \mathrm{i} \psi_t \ = \  -\psi_{xx} - 2 |\psi|^2 \psi + \epsilon \ v(x) \psi 
\label{eq:GP}
\ee
where $v(x)$ denotes the  attractive double well potential 
\be
 v(x)\ =\ -\delta(x-L) - \delta(x+L), \label{eq:v-def}
\ee
with conserved Hamiltonian energy functional
\be
\cH_{\mathrm{GP}}[\psi,\bar\psi]\ =\ \int\ \left(\left|\frac{\mathrm{d}\psi(x)}{\mathrm{d} x}\right|^2
 +v(x)|\psi(x)|^2-\frac{1}{2}|\psi(x)|^4\ \right)\ \mathrm{d} x
\label{eq:GPenergy}
\ee
and conserved particle (boson) number
\be
\cN[\psi,\bar\psi]\ =\ \int |\psi(x)|^2\ \mathrm{d} x.
\label{eq:Ndef}
\ee
Symmetry breaking bifurcations have been observed in this model
\cite{GMC} and in the analogous double square well model \cite{MKR},
by piecewise analytical solution in terms of hyperbolic secant and Jacobi 
elliptic function solutions of the translation invariant NLS equation ($v=0$),
followed by numerical solution of the nonlinear algebraic equations resulting
from the required continuity and jump conditions at singular points of the 
equation's coefficients.

\bigskip

We present a simple geometric analysis demonstrating
\begin{itemize}
\item[(A)] 
{\it the symmetry breaking of the ground state, as a bifurcation of an asymmetric branch 
of solutions from the branch of symmetric solutions}
\end{itemize}
and
\begin{itemize} 
\item[(B)] 
{\it the exchange of dynamical stability from the symmetric branch to the asymmetric branch. 
In particular, beyond the bifurcation point the symmetric state is linearly exponentially 
and nonlinearly unstable, whereas the asymmetric state is nonlinearly orbitally stable.}
\end{itemize}
We note that our analysis does not require the specific cubic nonlinearity
and is applicable for general local nonlinearities, where $|\psi|^2\psi$
is replaced by $f(|\psi|^2)\psi$, with $f(\cdot)$ of a general class.

\bigskip

\begin{figure}[htd]
\bigskip  
\begin{center}
\epsfxsize = 8.0cm \epsffile{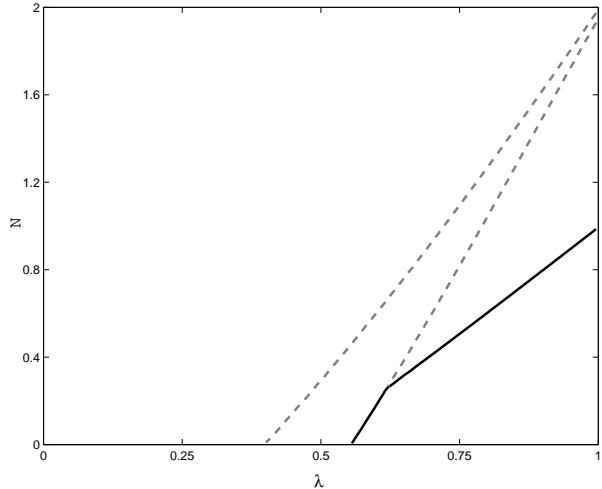}
\end{center}
\caption{
A bifurcation diagram is given for the case $\epsilon=1$, $L=2$.
The curve on the left bifurcating from $\cN=0$ represents the antisymmetric
pulse, whereas the curve on the right represents the symmetric pulse.  At
the threshold value $\cN_{\rm thresh}$, an asymmetric pulse bifurcates from 
the symmetric pulse.  At $\cN-$ levels where the asymmetric pulse is present,
the symmetric pulse is unstable.}
\label{fig:bif_diagram}
\bigskip  
\end{figure}

A bifurcation diagram showing the symmetric, antisymmetric and asymmetric
states is presented in figure \ref{fig:bif_diagram}. The solid curve
corresponds to the nonlinearly stable ground state. As $\lambda$ is increased
($-\lambda^2$ is the frequency) the symmetric solution becomes unstable
at the bifurcation point, while the new asymmetric state is nonlinearly stable.
We expect that anti-symmetric states (nonlinear excited states), corresponding to 
points on the disconnected branch to the left of the symmetric branch, 
are unstable to generic perturbations due to a nonlinear resonance phenomena, 
and that for large $t$, the system's energy resides in the ground state and 
radiation modes; see, for example, \cite{SW:99,SW:RMP}. 

\section{Overview of Gross-Pitaevskii / NLS}\label{sec:overviewGPNLS}
\bigskip

In appendix \ref{sec:wellposedness} we review the well-posedness theory of
(\ref{eq:GP},\ref{eq:v-def}). 
The initial value problem is well-posed in $C^0(R^1;H^1(R^1))$.

In this section we focus on  {\it nonlinear bound states} of (\ref{eq:GP}). 
These are solutions of the form:
\be 
\psi_b(x,t)\ =\ \mathrm{e}^{\mathrm{i} \lambda^2 t} \psi(x;\lambda),\  \psi \in L^2.
\label{eq:ub}
\ee
For the linear Schr\"odinger equation, the functions $\psi$ are
eigenstates of a Schr\"odinger operator: $-\D_x^2+v(x)$
and satisfy the equation
\be
-\psi_{xx}\ +\ v(x)\psi =
 -\lambda^2\psi .
\label{eq:linear-efn}
\ee
In the case where $v(x)$ is given by (\ref{eq:v-def}), with $L$ sufficiently
large, the set of bound states is two-dimensional, spanned by symmetric and 
antisymmetric functions. 

For (\ref{eq:GP}), nonlinear bound states satisfy the equation
\be
-\psi_{xx}\  +\ \left(\ v(x)\  -\
  2|\psi|^2\ \right)\psi\  =\
 -\lambda^2\psi.
\label{eq:varphi-eqn}
\ee
For the translation-invariant case, $v=0$, there is a well-known 
family of solitary (``soliton'') traveling wave solutions.
These are Galilean boosts (see (\ref{eq:Galilei})) of the basic solitary
(``soliton'') standing wave with hyperbolic secant spatial profile.
For a non-trivial potential $v(x)$ as in (\ref{eq:v-def}), the equation
is no longer translation-invariant and we have ``defect'' or
``pinned'' states. These pinned nonlinear bound states are plotted in the 
bifurcation diagram of figure \ref{fig:bif_diagram}.  In the translation 
invariant case, $v\equiv0$, the nonlinear bound states reduce to the branch 
of NLS solitons, bifurcating from the zero state at zero frequency,  
the endpoint of the continuous spectrum of the linearized operator
$-\D_x^2$ about the zero state.  For $v(x)$ nontrivial, there is a family of 
defect states, bifurcating from the zero state at each linear eigenfrequency 
of the linearized operator,  $-\D_x^2 + v(x)$ and in the direction of the 
corresponding eigenfunction; see \cite{RoseWeinstein:88} for a general discussion. \\

\noindent {\it Stability of nonlinear bound states:}\
An important  characterization of the nonlinear bound states is variational. 
The advantage of the variational characterization is that it can be used to 
establish nonlinear stability of the ground state;
see \cite{Weinstein:86,RoseWeinstein:88}. 

\begin{thm}
\label{thm:variational}

\noindent (I)\ Nonlinear bound states can be constructed  
variationally by  minimizing the Hamiltonian, $\cH$, subject to
fixed $L^2$-norm, $\cN$:
\be
\min_\varphi\  \cH[\varphi],\ \  \cN[\varphi]=\rho.
\label{eq:var}
\ee
Such minimizers are called {\it nonlinear ground states},
$\psi_{\rm Def}(\rho)$. 
The associated frequency, $-\lambda^2(\rho)$,
arises as a Lagrange multiplier for the constrained
variational problem \eqref{eq:var}. As $\rho\to0$, $-\lambda^2(\rho)$ 
tends to the ground state eigenvalue of $-\D_x^2+v(x)$.

\noindent (II)\ Ground states are $H^1$
nonlinearly orbitally Lyapunov stable, {\it i.e.}, if the initial 
data are $H^1$ close to a soliton (modulo the phase and translation 
symmetries of GP / NLS), then the solution remains close to a soliton 
in this sense for all $t \in (-\infty,\infty)$.
\end{thm}
\medskip

Although, the above theorem establishes the ground state as a 
dynamically stable state, it leaves unaddressed the question of detailed 
spatial characteristics of the ground state. As we shall now see, dynamical 
systems methods can, in some cases, be used to study this question 
globally and in detail.

\section{Nonlinear bound states - dynamical systems approach}

\noindent Dynamical systems methods are particularly well-suited to studying
the existence and nature of nonlinear bound states for the one-dimensional problem 
(\ref{eq:GP},\ref{eq:v-def}), since the equation is piecewise autonomous.
In particular, we study
\be
-\psi_{xx}   +\epsilon \  v(x) \psi  -2\psi^3\ =\ -\lambda^2\psi
\label{eq:sw-ode}
\ee
where $v(x)=-\delta(x+L)-\delta(x-L)$.
With a quick apology for the following non-physical substitutions, 
we rescale the variables as follows:
\ba
x & = & z / \lambda \nn\\
\psi(x) & = & \lambda \ \phi(\lambda x) = \lambda \ \phi(z).
\ea
In these variables, we have
$$
\phi_{zz} = \phi - 2 \ \phi^3 - \frac{\epsilon}{\lambda^2} \ v(z/\lambda) \phi.
$$
This is equivalent to the following system with matching conditions:
$$
\begin{array}{ll}
z \neq \pm \lambda L \qquad  & \ \ \ \ \phi^{\prime \prime} = \phi - 2 \ \phi^3         \\
& \\
z   =  \pm \lambda L \qquad  & \left\{  
\begin{array}{l}
 \phi(z^+) = \phi(z^-) \\
 \phi^{\prime}(z^+) = \phi^{\prime}(z^-)  - \frac{\epsilon}{\lambda} \ \phi(z), \\
\end{array}
\right.
\end{array}
$$
where $^{\prime}$ represents differentiation with respect to $z$ and 
 $f(z^\pm)=\lim_{\rho\to0}f(z\pm\rho)$.
As usual, we can convert a second order differential equation into
a pair of first order equations.  In particular, if we let
$(u,v) = (\phi, \phi^{\prime})$, then we have
$$
\begin{array}{ll}
z \neq \pm \lambda L \qquad  & \left\{
\begin{array}{l}
 u^{\prime} = v \\
 v^{\prime} = u  - 2 u^3 \\
\end{array}
\right. \\
& \\
z = \pm \lambda L \qquad  & \left\{
\begin{array}{l}
 u(z^+) = u(z^-) \\
 v(z^+) = v(z^-) - \frac{\epsilon}{\lambda} u(z). \\
\end{array}
\right.
\end{array}
$$

\noindent A search for bound states (standing waves) in this system can then be 
undertaken as follows:  
\begin{enumerate}
\item[1.]  Since any standing wave must decay as $z \rightarrow -\infty$,
a potential standing wave solution must originate (in the phase plane) 
along the global unstable manifold of $(0,0)$.  Without loss of generality, 
we consider only the portion of this manifold that lies in the first quadrant.

\item[2.]  When this solution approaches the first well at $z=-\lambda L^-$, 
we may shift $z$ so that the point 
$\lim_{z \rightarrow -\lambda L^-} \left(u(z),v(z)\right)$ can 
be chosen anywhere along the unstable manifold.  
(The value of $u(-\lambda L)$ can be thought of as a shooting parameter.)  
At this chosen point, the solution jumps vertically according to the matching 
conditions for $v$.

\item[3.]  After completing the vertical jump at $z=-\lambda L$, we again 
flow the solution forward for ``time'' $2 \lambda L$ 
according to the ordinary differential equation, until we approach
the second well at $z= +\lambda L$.

\item[4.]  When $z=+\lambda L$, the solution again jumps vertically
in the phase plane according to the matching condition.  The phase plane 
solution that we have been following represents a standing wave 
\emph{if and only if} this second jump lands our solution exactly 
upon the stable manifold, in accordance with the condition that 
$u(z) \rightarrow 0$ as $z \rightarrow +\infty$.

\end{enumerate}

\begin{figure}
\begin{center}
\begin{tabular}{cc}
\epsfxsize = 5.6cm \epsffile{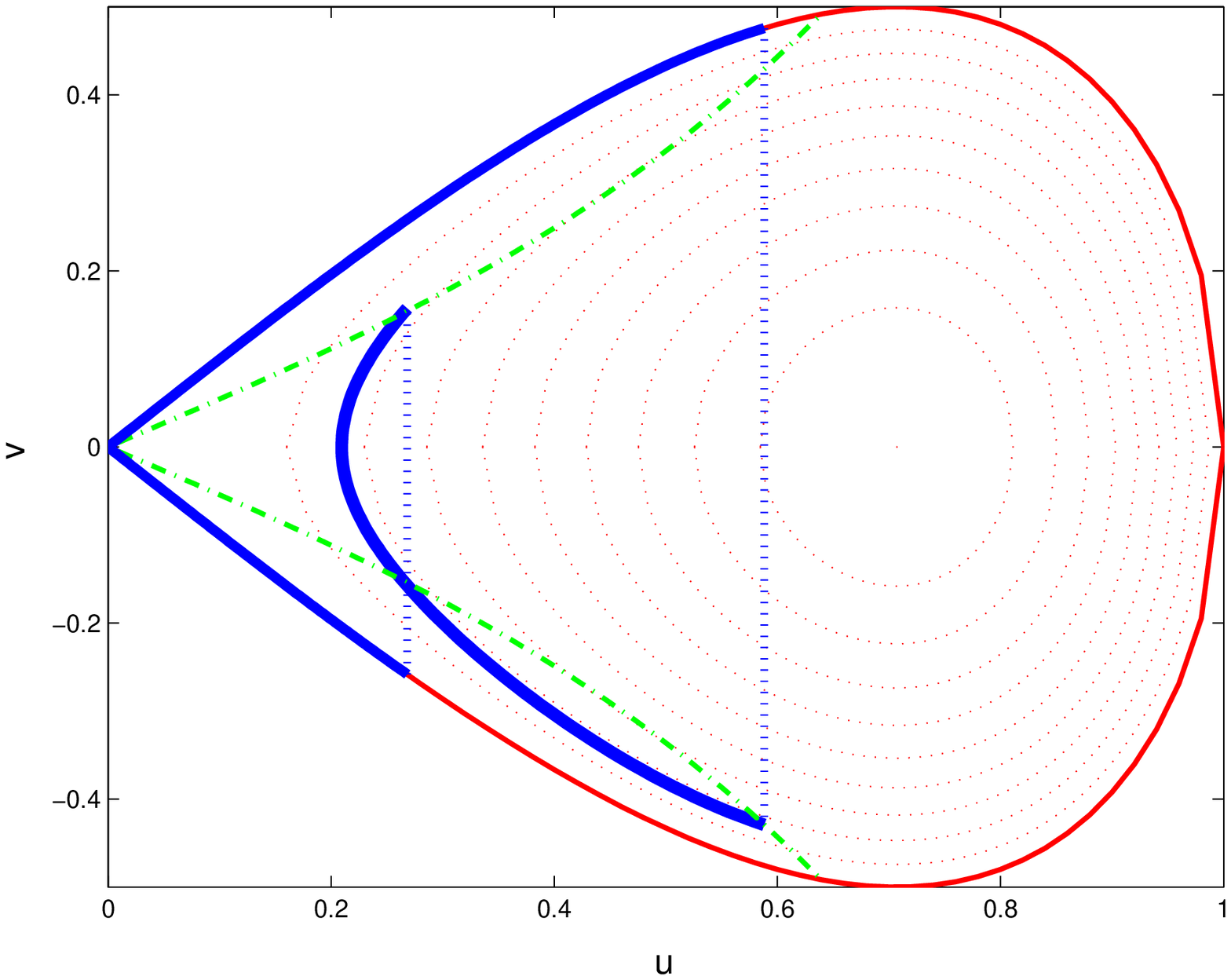} &
\epsfxsize = 5.6cm \epsffile{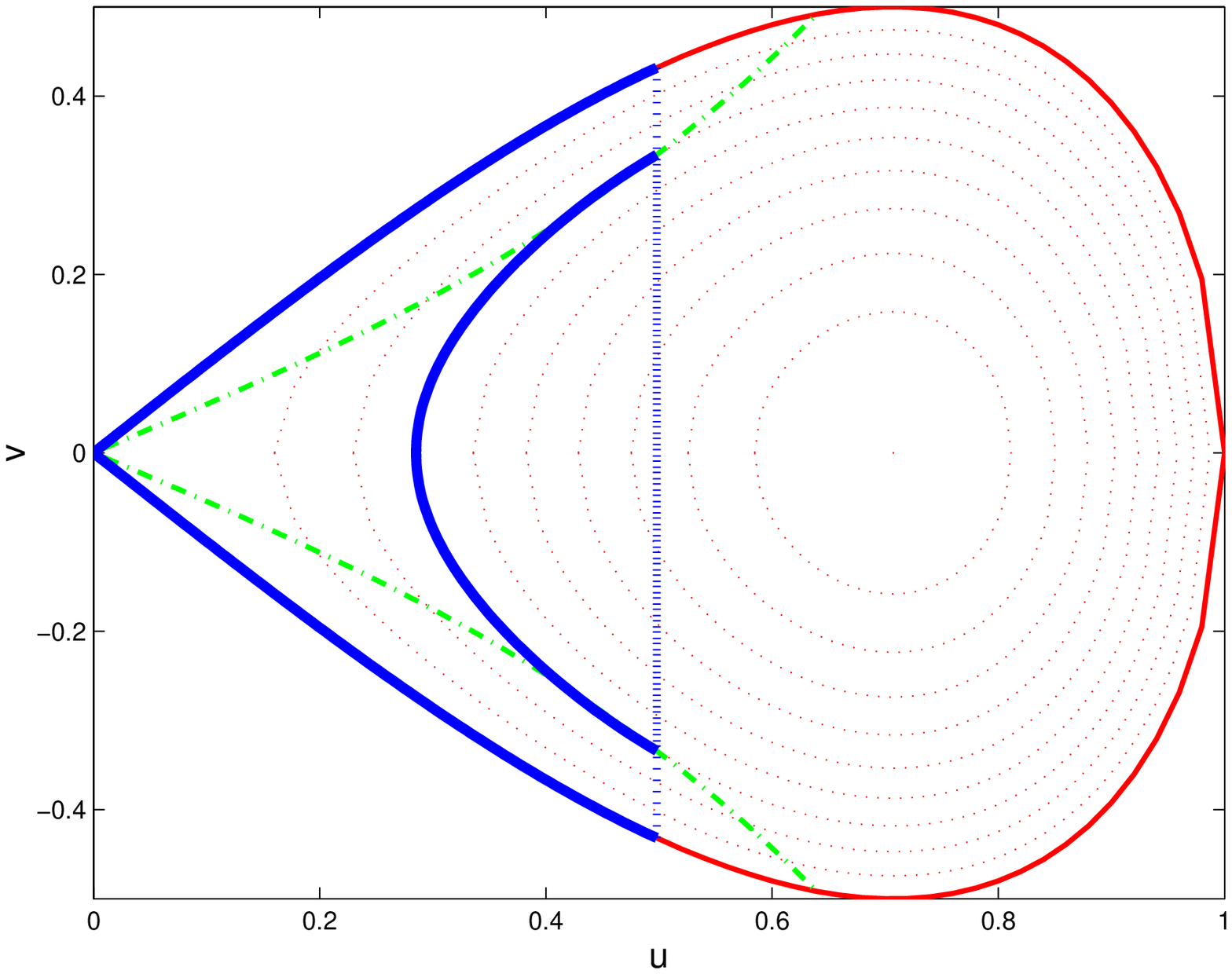} \\
\end{tabular}
\epsfxsize = 8.4cm \epsffile{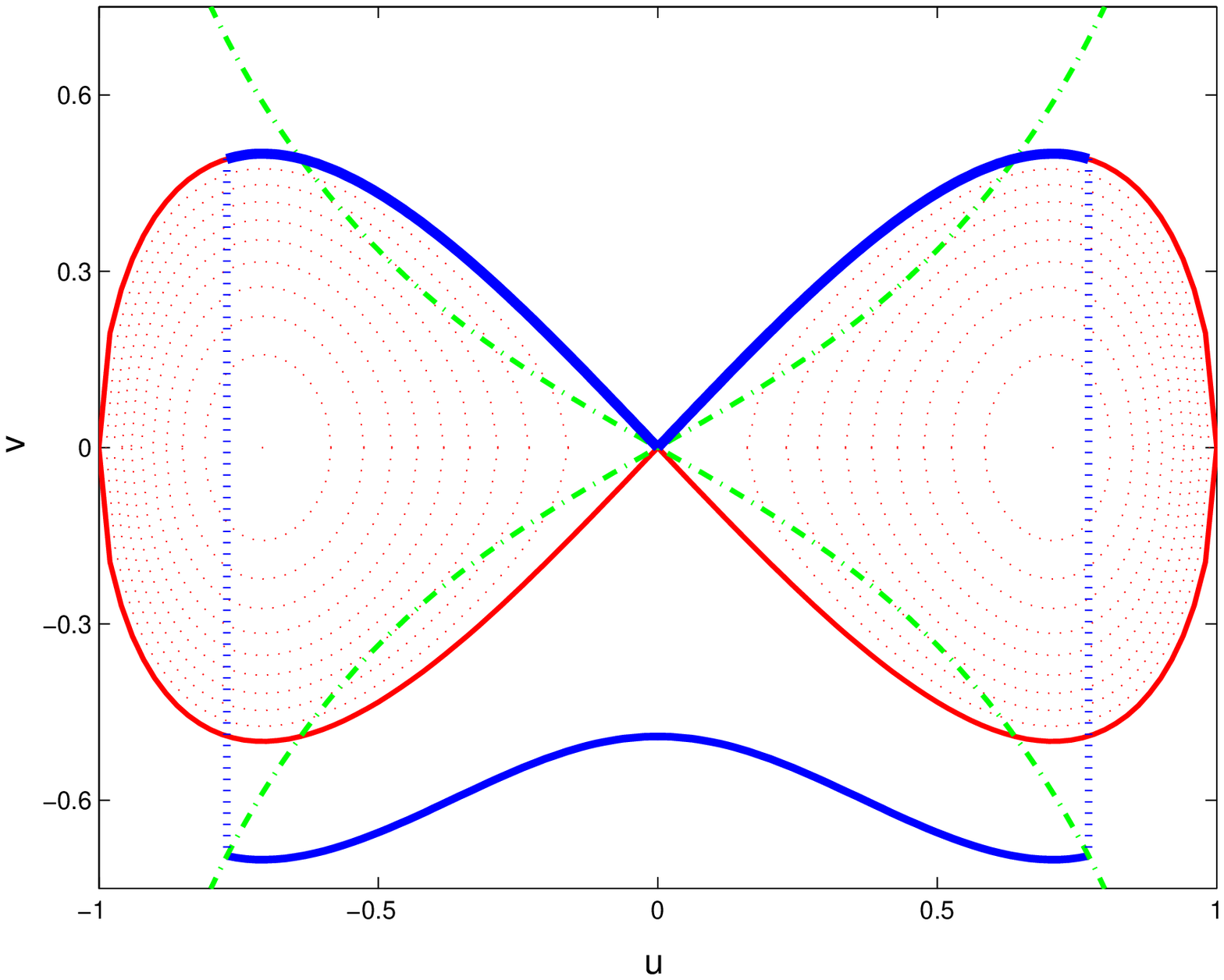}
\end{center}
\caption{Standing waves of different forms are found for 
$\epsilon=1$, $L=2$ and $\lambda=0.65$.  Here we see a symmetric wave,
an asymmetric wave and an antisymmetric wave.}
\label{fig:phase_plane}
\end{figure}

For fixed values of the parameters $\epsilon$, $\lambda$ and $L$, it may
happen that there are multiple standing waves present, corresponding to
different values of the function at the first defect.  In accordance with
the search method described above, such solutions are shown in the phase 
plane in figure \ref{fig:phase_plane}. 

\subsection{The linear limit; infinitesimal solutions}

To illuminate this search for standing waves, it is instructive to first 
consider the case of small amplitude solutions.  For infinitesimal 
solutions, we neglect the nonlinear term and we are left with the 
following \emph{linear} system

$$
\begin{array}{ll}
z \neq \pm \lambda L \qquad  & \left\{
\begin{array}{l}
 u^{\prime} = v \\
 v^{\prime} = u  \\
\end{array}
\right. \\
& \\
z = \pm \lambda L \qquad  & \left\{
\begin{array}{l}
 u(z^+) = u(z^-) \\
 v(z^+) = v(z^-) - \frac{\epsilon}{\lambda} u(z). \\
\end{array}
\right.
\end{array}
$$

For the linear system, a bounded solution again grows from $(0,0)$ along 
the unstable \emph{subspace}, jumps to some transient at the first well,  
evolves further according to the linear system and then must jump exactly
to the stable \emph{subspace} at the second well in order to decay to 
$(0,0)$ as $z \rightarrow \infty$.  For a fixed value of $L$ and 
$\epsilon$ in this system, we can determine the values of $\lambda$ 
for which standing waves exist.  This is pursued below.

\bigskip

\noindent
First, the unstable subspace at $(0,0)$ 
 is given as the eigenvector of the matrix
$$
\left( \begin{array}{cc} 0 & 1 \\ 1 & 0 \\ \end{array} \right)
$$
associated with the eigenvalue of positive real part.
This is the vector $(1,1)$ and so any point in the unstable subspace
can be written as $(\phi, \phi)$.  Similarly, the stable subspace
is spanned by the vector $(1,-1)$.  The jump at the first well is
then $J:(\phi, \phi) \rightarrow 
\left( \phi, (1-\frac{\epsilon}{\lambda})\phi \right)$.  
Note that this jump will take us to one of the \emph{interior} 
transient curves (inside the homoclinic loop in phase space) when 
$ \lambda > \epsilon / 2$ and to one of the \emph{exterior} transient 
curves (outside the homoclinic loop) when $\lambda < \epsilon / 2$.

\begin{figure}[htd]
\begin{center}
\epsfxsize = 8.0cm \epsffile{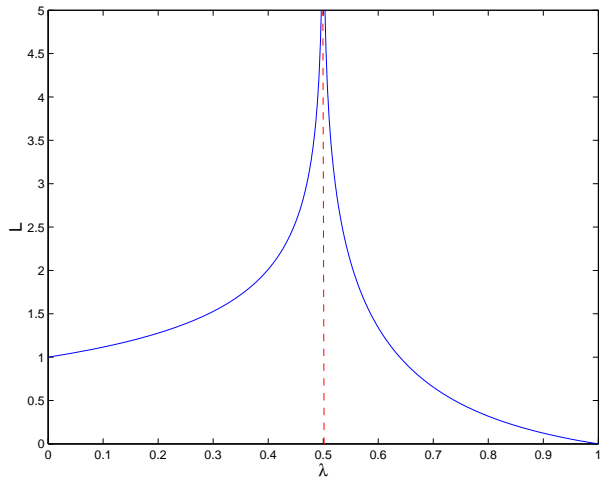} 
\end{center}
\caption{$\lambda$ vs. $L$ for $\epsilon = 1$. Symmetric states exist for any 
 $\lambda>\epsilon/2$, while for $\lambda<\epsilon/2$, antisymmetric states exist 
 only for well-separation $L>\epsilon^{-1}$.}
\label{fig:laml}
\end{figure}

\noindent
In this infinitesimal amplitude limit, we can find the standing waves of this system
explicitly.  Writing the landing point of the first jump in terms of the 
eigenbasis yields
$$
\left( \phi, \left(1-\frac{\epsilon}{\lambda}\right) \phi \right) 
 = \phi \left(1-\frac{\epsilon}{2 \lambda}\right) (1,1) 
    + \phi \frac{\epsilon}{2 \lambda} (1,-1).
$$
Because the system is linear in this limit, we ignore the $\phi$-terms
from here on.  Evolving this point forward for time $2 \lambda L$ brings
us to the point 
\ba
T_{2 \lambda L}
 \left(\left(1, 1-\frac{\epsilon}{\lambda}\right) \right) 
& = & 
T_{2 \lambda L}
\left( \left[1-\frac{\epsilon}{2 \lambda}\right] (1,1) 
    + \frac{\epsilon}{2 \lambda} (1,-1) \right)\nn \\
& = & 
\left[1-\frac{\epsilon}{2 \lambda}\right]\mathrm{e}^{2 \lambda L} (1,1) 
    + \frac{\epsilon}{2 \lambda} \mathrm{e}^{-2 \lambda L}(1,-1)\nn \\
& = & \left(
 \mathrm{e}^{2 \lambda L} 
 - \frac{\epsilon}{2 \lambda}\left[\mathrm{e}^{2 \lambda L} - \mathrm{e}^{- 2 \lambda L}\right],
 \mathrm{e}^{2 \lambda L} 
 - \frac{\epsilon}{2 \lambda}\left[\mathrm{e}^{2 \lambda L} + \mathrm{e}^{- 2 \lambda L}\right]
       \right).\nn\\
&&\label{eq:thatpt}
\ea
Now, the solution we are tracing will represent a standing wave
if and only if this point is taken to the stable subspace of
zero at the second jump.  By the jump conditions at $z=\pm\lambda L$ 
 and the fact that points on the unstable manifold are of the form $(\phi,-\phi)$,
 such points have the form
\be
J^{-1}(W^s) = 
  \left( \phi, \left(\frac{\epsilon}{\lambda} - 1 \right) \phi \right).
\label{eq:linJm1Ws}
\ee
Checking to see if the point (\ref{eq:thatpt})  is in this form yields
the following equation
$$
\left( \epsilon - \lambda \right)
 \left( 
 \mathrm{e}^{2 \lambda L} 
 - \frac{\epsilon}{2 \lambda}\left(\mathrm{e}^{2 \lambda L} - \mathrm{e}^{- 2 \lambda L}\right)
 \right)
 - \lambda  
 \left( 
 \mathrm{e}^{2 \lambda L} 
 - \frac{\epsilon}{2 \lambda}\left(\mathrm{e}^{2 \lambda L} + \mathrm{e}^{- 2 \lambda L}\right)
 \right) = 0.
$$
Some algebraic manipulation shows that this equation is equivalent to
$$
L =  \frac{1}{2 \lambda} \ln \left(\frac{ \pm \epsilon}{2 \lambda - \epsilon} \right).
$$
It may be more natural to consider $\lambda$ as a function of $L$
rather than vice versa; this would allow us to determine for which
values of $\lambda$ (frequencies $-\lambda^2$) stationary waves exist.  
This can be given implicitly as
$$ 
 \lambda = \frac{\epsilon}{2}\left(1\pm \epsilon \mathrm{e}^{-2\lambda L}\right)
$$
In particular, for $\lambda^2$ to be real and positive, 
we need the argument of the $\log$ to be  positive and 
therefore we have the following two solutions
\ba
&&0 < \lambda < \epsilon/2\ \   \ \ \ \ 
 L_{\rm anti} = \frac{1}{2 \lambda} \ln 
 \left(\frac{\epsilon}{\epsilon - 2 \lambda} \right)\label{eq:Lanti}\\
&&\epsilon/2 < \lambda < \epsilon\  \ \ \ \ 
 L_{\rm sym} = \frac{1}{2 \lambda} \ln 
 \left(\frac{\epsilon}{2 \lambda - \epsilon} \right). \label{eq:Lsym}
\ea
The antisymmetric solutions exist only for $L>\epsilon^{-1}$ because the function
$L_{\rm anti}(\lambda)$ is monotone increasing on 
 $[0,\epsilon/2)$, with minimum
value $\epsilon^{-1}$ attained at $\lambda=0$.

\noindent
The two branches of solutions (symmetric and antisymmetric)
are shown in figure \ref{fig:laml}. The results are summarized in 
the following proposition:

\begin{prop}
In the small amplitude limit, with the wells separated by any
positive distance $L > 1/\epsilon$, there exists a symmetric standing 
wave with $\epsilon/2 < \lambda(L) < \epsilon$ and an antisymmetric wave 
with $0 < \lambda(L) < \epsilon/2$.  These states are unique 
(up to linear scaling) and no other standing waves exist.
Finally, for $\epsilon L$ sufficiently large,
\be \lambda_\pm \sim \frac{\epsilon}{2}\left(1\pm \mathrm{e}^{-\epsilon L}\right),\
\     \lambda_+ - \lambda_- \sim\  \epsilon \mathrm{e}^{-\epsilon L}
\nn\ee
\end{prop} 

\subsection{Nonlinear regime; solutions with larger amplitude.}

For solutions with nontrivial amplitude, things get much more interesting
(as we have seen in our initial numerical search).
Because of the periodicity of the transient states, there are many possible 
solutions for this system (especially for larger values of $L$). 
We show particularly that in addition to the symmetric pulse (and the 
antisymmetric pulse, for $L$ values at which it exists) described above, 
there is also an asymmetric pulse that bifurcates away from the symmetric 
pulse at a particular value of the boson number $\cN$. 
(These results follow for a focusing nonlinearity, $g<0$.  In the defocusing
case, $g<0$, an asymmetric pulse can be seen to bifurcate from the 
{\it antisymmetric solution}.)

\bigskip

\noindent
For fixed values of the parameters $L$ and $\epsilon$, we will
consider what sorts of positive-valued standing waves exist for
varying values of $\lambda$.  These solutions can be determined by 
adapting the previous search technique in the following way:

\noindent 
\begin{enumerate}
\item[1.]  For this system, the branch of the unstable manifold in the 
first quadrant is given by the explicit expression
\be
W^u = \left\{ (u,v) : v = u \sqrt{1-u^2},\ 0\le u\le1 \right\}.
\label{eq:Wu}
\ee
We now search for standing waves by evolving this {\it entire curve} via
the matching conditions and the evolution equation.

\item[2.]  For a particular value of $\lambda$, application of the 
matching conditions at the first defect yields the curve
\be
J(W^u) = \left\{ (u,v) : v = u \left( \sqrt{1-u^2}
            - \frac{\epsilon}{\lambda} \right) \right \};
\label{eq:JWu}
\ee
see the dashed curve in figure \ref{fig:phase_plane}.
\item[3.]  We then evolve this curve for time $2 \lambda L$, yielding 
the curve
$$
T_{2 \lambda L}\left(J(W^u)\right).
$$

\item[4.] Finally, we compare the evolved curve 
$T_{2 \lambda L}\left(J(W^u)\right)$ with the set of points
that jump to the stable manifold at the second defect
\be
J^{-1}(W^s) = \left\{ (u,v) : v = - u \left( \sqrt{1-u^2}
            - \frac{\epsilon}{\lambda} \right) \right \}.
\label{eq:Jm1Ws}
\ee
This agrees with (\ref{eq:linJm1Ws}) in the linear limit.
Notice that the set $J^{-1}(W^s)$ is simply the reflection
of the set $J(W^u)$ through the $v$-axis.  Each non-trivial point of 
intersection of the two sets $T_{2 \lambda L}\left(J(W^u)\right)$
and $J^{-1}(W^s)$ represents a standing wave of this system.
\end{enumerate}

\bigskip

\begin{rem}
Because the curve $J(W^u)$ intersects many of the periodic
transients more than once, it seems plausible that there may be 
non-symmetric solutions for some values of $L$.  We will make
that clear by pursuing the strategy above.
\end{rem}

\begin{prop} Let $\epsilon$ and $L$ remain fixed.  For all 
$\lambda > \lambda_{\rm sym}$ (where $\lambda_{\rm sym}$ is the 
value of $\lambda$ for which a symmetric solution exists in
the linear limit, as in (\ref{eq:Lsym})), there is at least one 
symmetric solution standing wave. 
\end{prop}

\begin{figure}
\begin{center}
\begin{tabular}{cc}
\epsfxsize = 5.6cm \epsffile{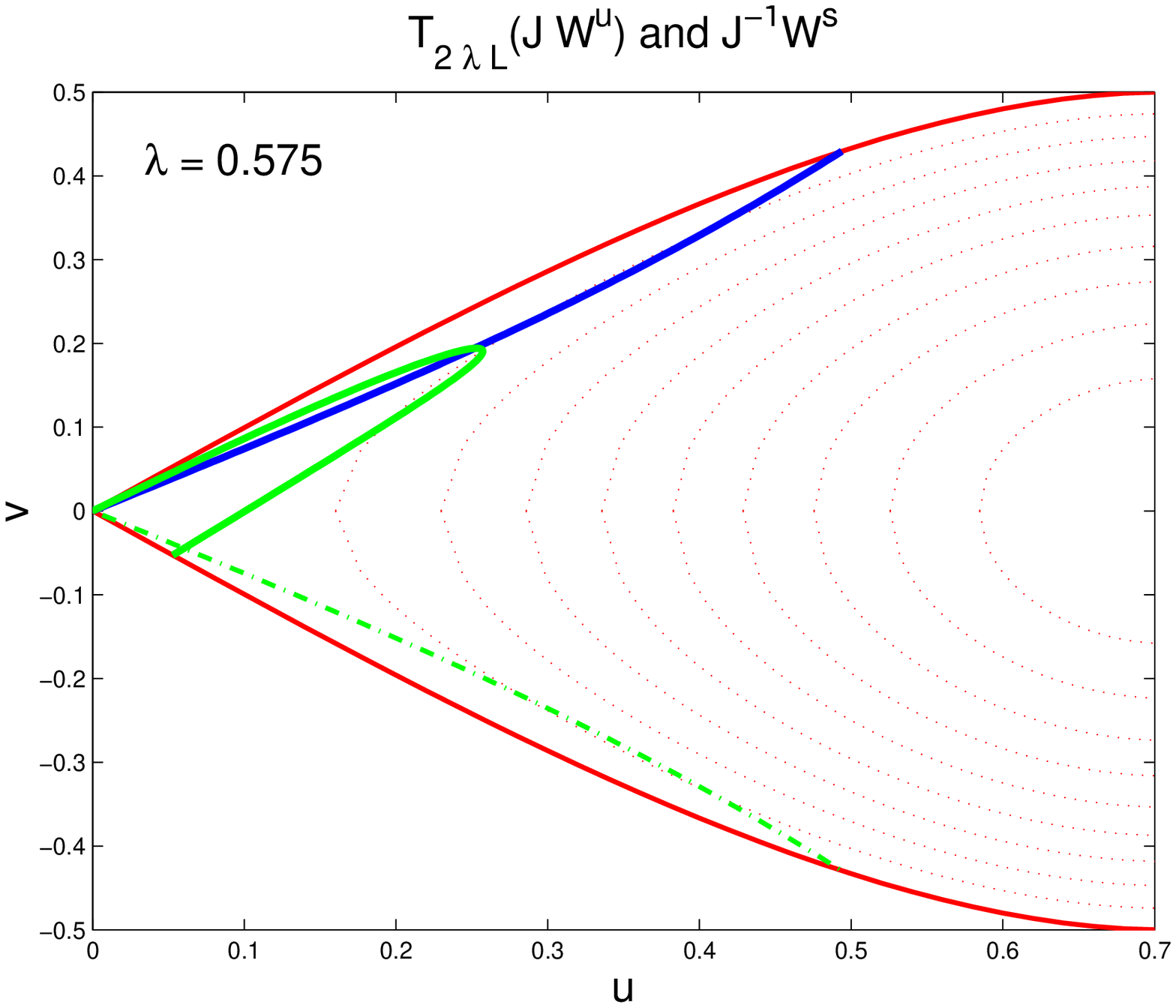} &
\epsfxsize = 5.6cm \epsffile{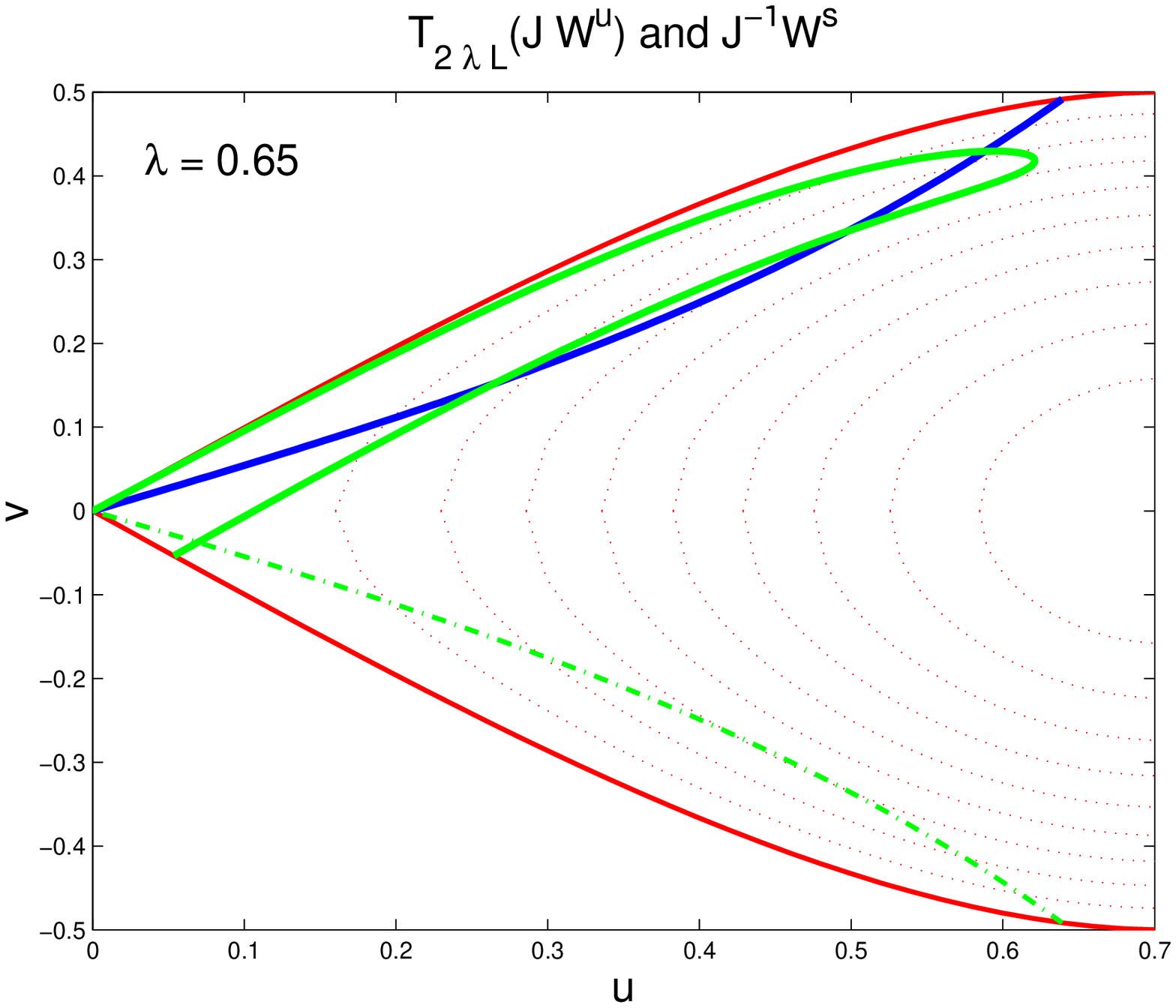} \\
\end{tabular}
\end{center}
\caption{For all values of $\lambda > \lambda_{\rm sym}$ there is an intersection
of $T_{2 \lambda L}\left(J(W^u)\right)$ and $J^{-1}(W^s)$ -- and such an intersection
represents a standing wave.  As $\lambda$ increases, more intersections may arise.}
\label{fig:curves}
\end{figure}

\begin{proof}
This is a simple argument regarding the continuity of the curves
$J^{-1}(W^s)$ and $T_{2 \lambda L}\left(J(W^u)\right)$.  We consider
only the portions of these curves interior to the homoclinic orbit
of the autonomous system.  The portion of $J(W^u)$ inside this loop
has one endpoint at $(0,0)$ and the other endpoint on the
curve $W^s$ in the lower half-plane; see figure \ref{fig:curves}.  

For $\lambda > \lambda_{\rm sym}$, 
 the initial portion of the curve $T_{2 \lambda L}\left(J(W^u)\right)$
leaving $(0,0)$ is between $J^{-1}(W^s)$ and $W^u$ -- in particular, 
\emph{above} the curve $J^{-1}(W^s)$.  However, the other end must
remain on the invariant manifold $W^s$, and this manifold lies 
\emph{below} the curve $J^{-1}(W^s)$.  Therefore, since
$T_{2 \lambda L}\left(J(W^u)\right)$ is a continuous curve, it must
intersect the curve $J^{-1}(W^s)$ in at least one point.  Therefore
there is at least one standing wave in this system for all 
$\lambda > \lambda_{\rm sym}$.

Finally, notice that if there is \emph{only} one point of intersection, then 
the standing wave corresponding to this intersection point will be
a symmetric function of position. 
  This follows from the reversibility of the autonomous
portion of the evolution equation.
\end{proof}

Knowing that one or more standing waves exist, we now look into the forms 
that such waves take.  In the following, we will look for symmetric 
standing waves that are concentrated at the defects; in particular,
we hope to find profiles that achieve their maximum values at the two 
defects.  We take a slight change of view in the following, considering 
$\epsilon$ and $\lambda$ as fixed and asking for which values of the 
spacing parameter $L$ do such standing waves exist?

\begin{prop} Let $\epsilon$ and $\lambda$ be fixed parameters.  There
is a symmetric solution \emph{concentrated at the defects} in each of
the following regimes:
\begin{enumerate}
\item[1.] For $\frac{\epsilon}{2} < \lambda < \epsilon$, there is such a 
solution for all $L > L_{\rm sym}(\lambda)=\frac{1}{2 \lambda} \ln 
 \left(\frac{\epsilon}{2 \lambda - \epsilon} \right)$.
\item[2.] For $\epsilon < \lambda < \sqrt{2} \epsilon$, there is such a 
solution for all values of $L$.
\item[3.] For $\lambda = \sqrt{2} \epsilon$, there is such a 
solution for all 
$L > \frac{1}{2 \epsilon}\left(\tan^{-1}2 \right)$.
\end{enumerate}
\end{prop}

\begin{proof}
We will prove this proposition by considering the amount of ``time'' it
takes each point on the curve $J(W^u)$ (from the point where the curve 
enters the lower half plane until it meets the manifold $W^s$) to reach 
its reflection on the curve $J^{-1}(W^s)$.  For fixed value $2L$, we show
there is a point on $J(W^u)$ which is mapped to its reflection
 on $J^{-1}(W^s)$ in time $2L$. We can define
the function $L(u)$ as the ``time'' it takes the solution beginning
at a point $\left(u, u(\sqrt{1-u^2} -\epsilon/\lambda)\right)$ (meeting the
previous parenthetical requirement) to reach the symmetric point
$\left(u, -u(\sqrt{1-u^2} -\epsilon/\lambda)\right)$.

\begin{enumerate}
\item[1.] 
The result follows quickly in the regime 
$\frac{\epsilon}{2} < \lambda < \epsilon$, by noting that as
$u \rightarrow 0^+$, $L(u) \rightarrow L_s$ as given in the earlier
proposition.  Also, as  $u \rightarrow 
\left(\sqrt{1 - (\epsilon/2\lambda)^2}\right)^-$,
$L(u) \rightarrow \infty$.  A continuity argument completes this case.

\item[2.]
 \ For $\lambda > \epsilon$ the curve $J(W^u)$ leaves the origin into
the upper half plane.  It reaches the $u$-axis when 
$u=\sqrt{1 - (\epsilon/\lambda)^2}$ and reaches the
stable manifold $W^s$ when $u=\sqrt{1 - (\epsilon/2\lambda)^2}$.
In this regime, it is clear that $L(u) \rightarrow 0$ as
$u \rightarrow \left(\sqrt{1 - (\epsilon/\lambda)^2}\right)^+$ and
$L(u) \rightarrow \infty$ as
$u \rightarrow \left(\sqrt{1 - (\epsilon/2\lambda)^2}\right)^-$

\item[3.] 
 When $\lambda = \sqrt{2} \epsilon$, the curve $J(W^u)$ crosses the
$u$-axis at the elliptic fixed point $(1/\sqrt{2},0)$.  A careful
study of the linearization near this fixed point reveals the lower
limit given above.
\end{enumerate}
\end{proof}

Now we note an important consideration concerning the form of these
symmetric pulses.  In particular, for fixed values of the
parameters $\epsilon$ and $\lambda$, there is a distinguished value
of the parameter $L$ so that the curve $J(W^u)$ is tangent to the
symmetric standing wave as we approach the first defect from the right.
The value of the standing wave at each defect can be quickly computed
for this threshold value,
$$
u_{\mathrm{thresh}} = \sqrt{\frac{12 -\left(\frac{\epsilon}{\lambda}\right)^2
 - \frac{\epsilon}{\lambda}\sqrt{12 + \left(\frac{\epsilon}{\lambda}\right)^2}}{18}}
$$
and the threshold value of $L$, $L_{\rm thresh}$, 
  can then be computed simply by
noting the ``time'' it takes to flow from 
$(u_{\mathrm{thresh}}, v_{\mathrm{thresh}})$ to its reflection
$(u_{\mathrm{thresh}}, -v_{\mathrm{thresh}})$.

\begin{figure}[htd]
\begin{center}
\epsfxsize = 8.0cm \epsffile{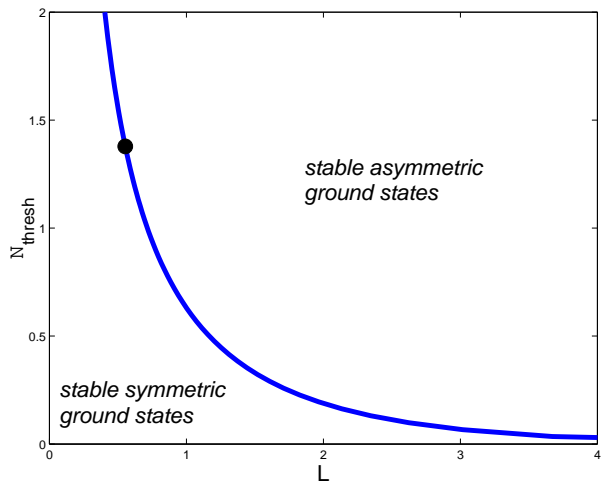} 
\end{center}
\caption{Threshold values of the particle number ${\cal N}_{\rm thresh}$
 for varying values of $L$ are shown.
At these threshold values, there is a bifurcation of asymmetric pulses 
(see Proposition \ref{prop:bif}) and beyond these values, the symmetric
standing wave will be unstable (see Theorem \ref{thm:stab}).  The point along the
curve marked with a solid circle corresponds to the special value 
$\lambda = \sqrt{2} \epsilon$, where the point $u_{\mathrm{thresh}}$ 
is the fixed point in the phase plane.  Along the curve to the right of this point, 
the symmetric standing waves at threshold have the desired two peak profile; to the left, 
the symmetric standing waves at threshold have a single peak centered between the two wells.}
\label{fig:threshold}
\end{figure}

In figure \ref{fig:threshold}, we show the energy of the
symmetric standing wave at the threshold values of $L$.  We will 
now show that beyond this threshold, there is a pair of {\it asymmetric 
standing waves} (in addition to the symmetric and antisymmetric waves 
discussed previously). 

\begin{prop}
\label{prop:bif}
 Let $\epsilon$ and $\lambda$ remain fixed. For $L > L_{\mathrm{thresh}}(\lambda)$,
there exists a pair of non-symmetric standing waves.  These waves leave the symmetric 
wave in a pitchfork bifurcation at $L = L_{\mathrm{thresh}}$.

\end{prop}

\begin{figure}
\begin{center}
\begin{tabular}{cc}
\epsfxsize = 5.6cm \epsffile{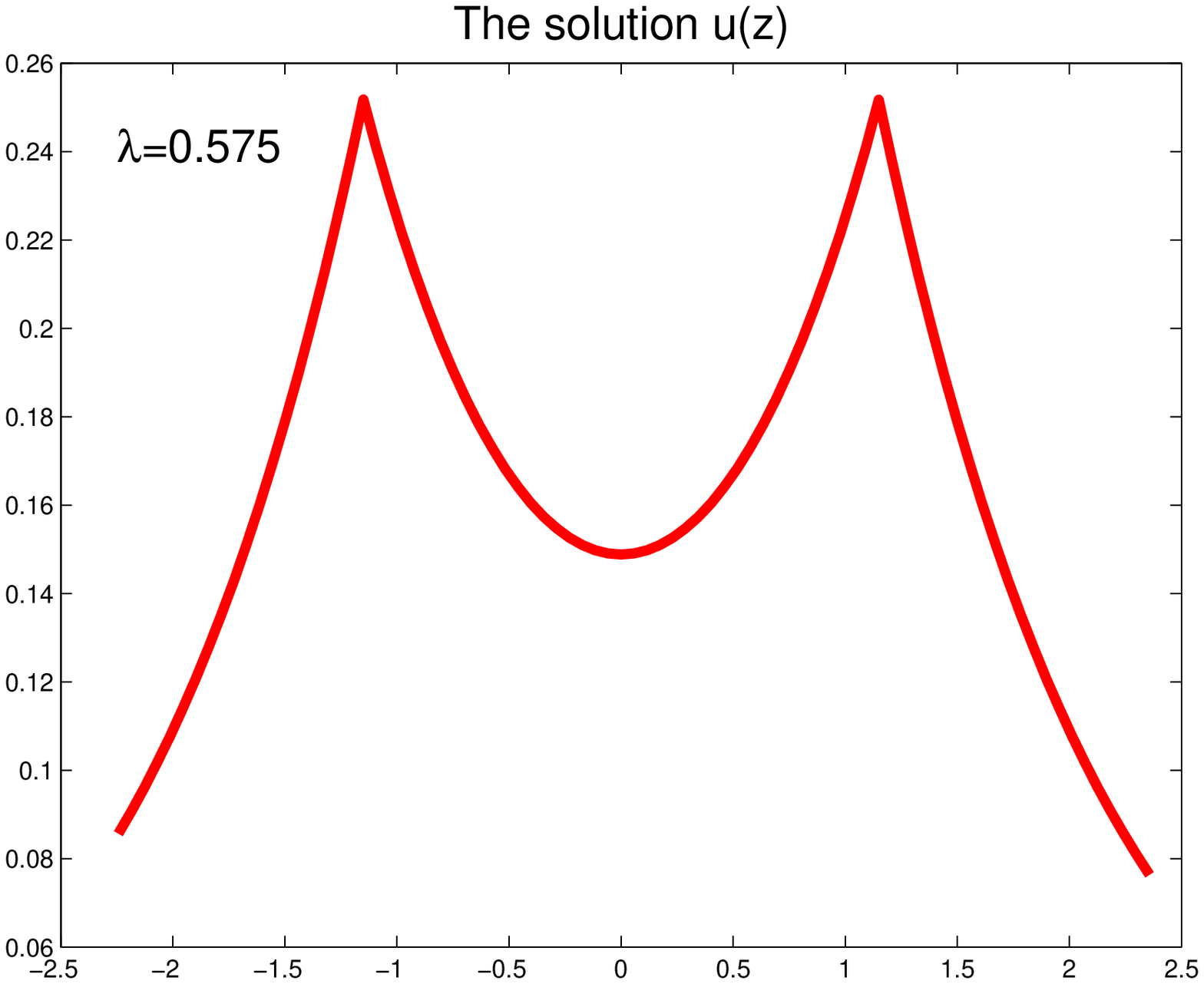} &
\epsfxsize = 5.6cm \epsffile{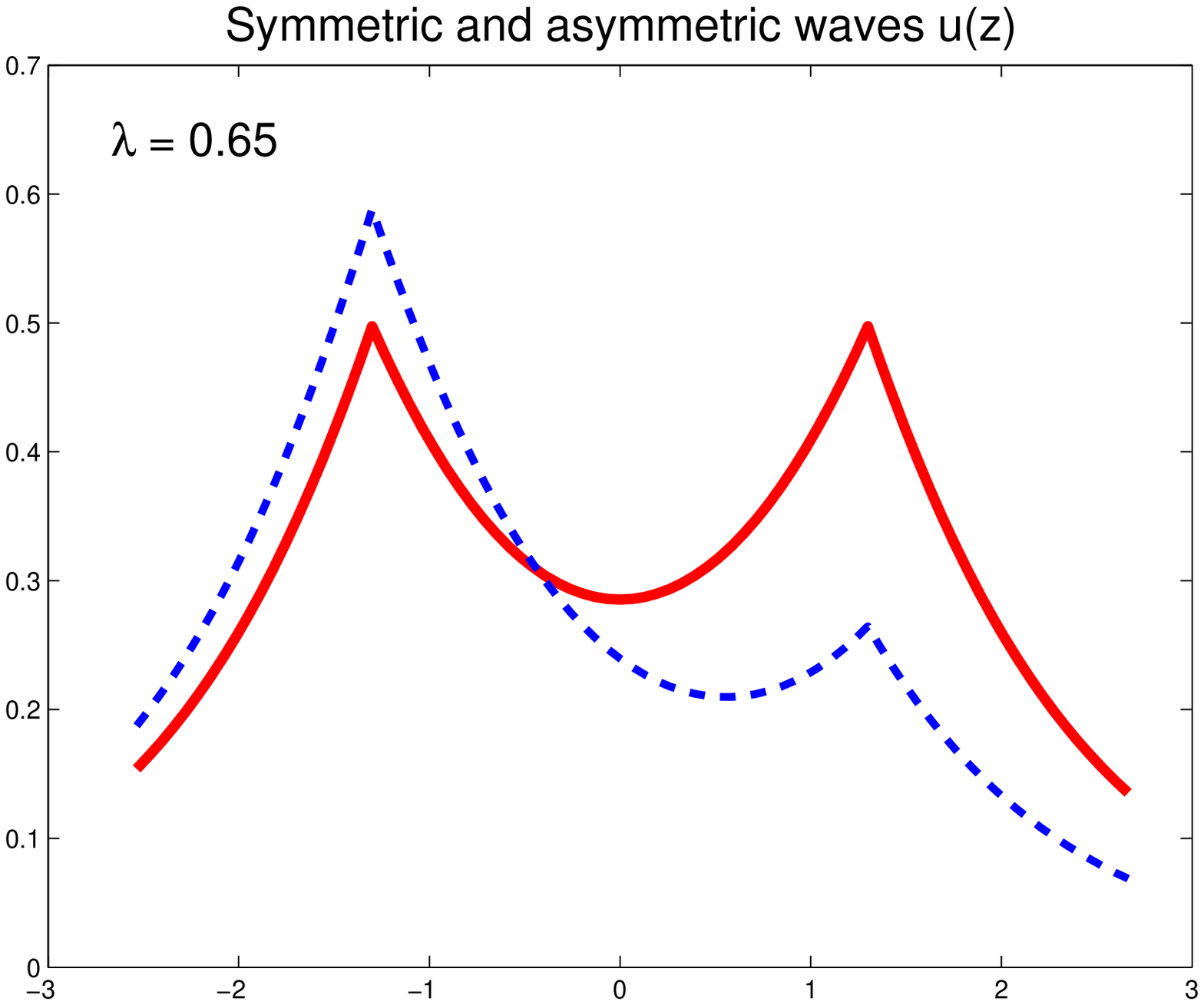} \\
\end{tabular}
\end{center}
\caption{The solutions corresponding to the various intersection points 
from the phase plane diagram in figure \ref{fig:curves} are shown here.}
\label{fig:solutions}
\end{figure}

\begin{proof}
Consider $\epsilon$ and $\lambda$ fixed and positive, with $L$ so that
$L > L_{\mathrm{thresh}}(\lambda)$.  For this value of $L$, the symmetric
standing wave has a value $(u_0,v_0)$, with $u_0 > u_{\mathrm{thresh}}$,
approaching the first defect from the right.  This wave then must pass 
through $J^{-1}(W^s)$ in the phase plane an additional 
time before the reaching the terminal point at the second defect.

In the phase space, the part of the standing wave between the two defects
can be written as the curve
$$
\Gamma_{0} = 
  \left\{ T_z\left((u_{0},v_{0})\right) : 
      0 \leq z \leq 2 \lambda L \right\}.
$$
Note that the curve $\Gamma_{0}$ simply traces out the portion of the 
periodic orbit connecting the symmetric points $(u_0,v_0)$ and $(u_0,-v_0)$.

We consider an arbitrary parameterization $(u_{\tau}, v_{\tau})$, 
$\tau \in [0,1]$ of the portion of the curve $J(W^u)$
between $(u_0,v_0)$ and its nontrivial intersection with the 
stable manifold $W^s$, so that $(u_0,v_0)$ is as above and 
$(u_1,v_1) \in W^s$.  Then we can define
$$
\Gamma_{\tau} = 
  \left\{ T_z\left((u_{\tau},v_{\tau})\right) : 
      0 \leq z \leq 2 \lambda L \right\},
$$
for all $\tau \in [0,1]$.  Notice that the curve $J^{-1}(W^s)$ is transverse
to all of the periodic orbits in the phase plane away from the fixed
point $(1/\sqrt{2},0)$.  Therefore if a curve $\Gamma_{\tau}$
crosses $J^{-1}(W^s)$, it must do so transversely.  

Recall that the curve $\Gamma_{0}$ crosses $J^{-1}(W^s)$ prior to 
its terminal point at the second defect.  (This earlier crossing
does not represent a standing wave, of course, since it occurs for
a value of $z$ not corresponding to a defect.)  On the other hand, for
$\tau$ sufficiently close to $1$, the curve $\Gamma_{\tau}$ no 
longer crosses $J^{-1}(W^s)$.  Continuity considerations therefore 
dictate that there must be a value of $\tau$ between $0$ and $1$ so
that $T_{2 \lambda L}\left( (u_{\tau},v_{\tau})\right)$ lies on the
curve $J^{-1}(W^s)$.  This value clearly does not represent a symmetric 
state since $u_0 > u_{\rm thresh}$, while the first point of intersection
is smaller than $u_{\rm thresh}$.

There is a second non-symmetric standing wave that, in the phase plane,
has the form of the wave just found reflected across the $u$-axis.  

This pair of non-symmetric solutions is originally produced in a pitchfork
bifurcation when $L = L_{\rm thresh}$ -- at this parameter value the 
transversality of $J^{-1}(W^s)$ and the periodic orbit carrying the 
symmetric standing wave is lost.

\end{proof}

\section{The eigenvalue problem and stability.}

In the previous section, we found various standing wave 
solutions of the equation
$$
\mathrm{i}\psi_t = -\psi_{xx} - 2 |\psi|^2 \psi + \epsilon \ v(x) \psi
$$
where $v(x)$ is the double well potential   
$-\delta(x-L) - \delta(x+L)$.
In this section, we will analyze the stability of the various
branches of solutions constructed, by studying the eigenvalue equation, 
 associated with linearization about a fixed nonlinear bound state.

First we change coordinates into the frame of the standing wave
$$
\psi(x,t) = \mathrm{e}^{\mathrm{i} \lambda^2 t} \phi(x,t)
$$
and arrive at the equivalent equation
\begin{equation}
\label{thepde}
\mathrm{i}\phi_t = - \phi_{xx} 
+ \left[\lambda^2 - \left(2 |\phi|^2 + \epsilon \ v(x)\right)\right] \phi.
\end{equation}
We write this equation in real and imaginary parts ($\phi=v+\mathrm{i} w$) 
\begin{equation}
\label{reimparts}
\begin{array}{ccccccc}
v_t & = &-w_{xx}& - &[2 (v^2 + w^2) + \epsilon \ v(x)] w& + &\lambda^2 w, \\
w_t & = & v_{xx}& + &[2 (v^2 + w^2) + \epsilon \ v(x)] v& - &\lambda^2 v \\
\end{array}
\end{equation}
and formally consider small perturbations of the standing wave.  Since
the standing wave (let's call it $u(x)$) is real-valued, we then write
$\left(v(x,t),w(x,t)\right) = \left(u(x),0\right) + 
\delta\left(p(x,t),q(x,t)\right)$ with $\delta$  small. Leading order in 
 $\delta$, yields the linearized dynamics 
$$
\begin{array}{rcrclcr}
p_t & = &-q_{xx} &-& [2 u^2 + \epsilon \ v(x)] q           &+& \lambda^2 q, \\
q_t & = & p_{xx} &+& [2 u^2 + \epsilon \ v(x)] p - 4 u^2 p &-& \lambda^2 p. \\
\end{array}
$$
Setting
$$
\begin{array}{rcl}
L_- & = & \frac{\mathrm{d}^2}{\mathrm{d}x^2} + [2 u^2 + \epsilon \ v(x)]         - \lambda^2, \\
&&\\
L_+ & = & \frac{\mathrm{d}^2}{\mathrm{d}x^2} + [2 u^2 + \epsilon \ v(x)] + 4 u^2 - \lambda^2, \\
\end{array}
$$
equation (\ref{reimparts}) can be rewritten 
 as the system 
$$
\left( \begin{array}{c} p \\ q \\ \end{array} \right)_t 
=  \left( \begin{array}{cc} 0 & -L_- \\ L_+ & 0 \\ \end{array} \right)
\left( \begin{array}{c} p \\ q \\ \end{array} \right)
= N \left( \begin{array}{c} p \\ q \\ \end{array} \right).
$$
Of interest is the spectrum of this operator $N$. For the ground state
 (see section \ref{sec:overviewGPNLS}), the spectrum of $N$ lies entirely
on the imaginary axis and modulo zero modes, associated with symmetries,
  $\exp(tN)$ is bounded on $H^1$ and indeed, the solution is nonlinearly stable
\cite{Weinstein:85,RoseWeinstein:88}.

 A state is 
unstable if one can exhibit an  
eigenvalue of the operator $N$ with positive real part. Nonlinear
 instability is a consequence; see the appendix of \cite{grillakis}.
Unfortunately, even though the spectrum of $L_+$ and $L_-$ can, 
in principle, be determined using Sturm--Liouville theory, it is not 
obvious how to construct the spectrum of $N$ using this information. 
However, this is exactly the operator studied by Jones in \cite{jones:88}, 
in an application to optical pulses in nonlinear waveguides.  
In that work, Jones interprets the search for a real positive eigenvalue 
as a shooting problem in the space of Lagrangian planes.  
Using a winding argument in this space, he derives a relationship between 
the number of positive eigenvalues of the simpler operators $L_+$ and $L_-$ 
and the number of positive eigenvalues of $N$. 
Let 
$$
\begin{array}{lcl}
P & = & \mbox{number of positive eigenvalues of $L_+$} \ \  {\rm and}\\
Q & = & \mbox{number of positive eigenvalues of $L_-$}. \\
\end{array}
$$
Then we have \cite{jones:88,grillakis}:
\begin{thm}\label{thm:jonesgrillakis}
If $P-Q \neq 0,1$, there is a real positive eigenvalue of the operator $N$.
\end{thm}
An alternative,
 PDE approach to this result was developed by Grillakis
\cite{grillakis}.

\bigskip

\noindent
From Sturm--Liouville theory, $P$ and $Q$ can be determined by 
considering the behavior of solutions of $L_+p = 0$ and $L_-q = 0$, 
respectively.
In particular, $P$ and $Q$ equal the number of nodes of the associated
solutions $p$ and $q$.  Since $L_-q = 0$ is satisfied by the 
standing wave $u(x)$ itself, it quickly follows that 
$$
Q = \mbox{number of zeros of the standing wave $u(x)$}.
$$
For all positive-valued pulses, then, $Q=0$.  

\bigskip

So we have that if $P > 1$, then $N$ has a positive eigenvalue and the 
standing wave $u(x)$ is unstable.  The operator $L_+$ acts as the 
equation of variations for (\ref{thepde}), and a natural interpretation
of the equation of variations is that it carries tangent vectors under the flow.
So if we initialize $q(x)$ as a vector tangent to the unstable manifold
at the point $u(x)$, then we can calculate $P$ by counting the nodes of 
the resulting solution $q(x)$.

In the context of vectors, a node of $q(x)$ is a vector in the phase plane
pointing straight up or down.  And so, $P$ can be computed by counting 
the number of times that the vector initialized tangent to the unstable 
manifold passes through verticality.  In particular, considering
only the angle of this vector and not its magnitude, $P$ corresponds
to the largest number of integer multiples of $\pi$ through which this
vector rotates as it is evolved under the linearized flow.

The only difficulty in the computation is this: because of the jumps at the defects, 
{\it i.e.}, the inhomogeneity of the evolution equation, a vector $q(x)$ initialized
tangent to the nonlinear bound state solution \emph{will not} remain tangent 
to this solution once we reach the first defect.

However, using the simple matching conditions for these vectors at the wells, 
we have the following theorem:

\begin{thm}
\label{thm:stab}
For $L \leq L_{\rm thresh}$, the symmetric solution has $P \leq 1$.  
For $L > L_{\rm thresh}$, the symmetric solution has $P \geq 2$.
Therefore, after the bifurcation of the asymmetric waves, 
the symmetric wave is unstable.
\end{thm}

\begin{figure}
\begin{center}
\begin{tabular}{c}
\epsfxsize = 6.7cm \epsffile{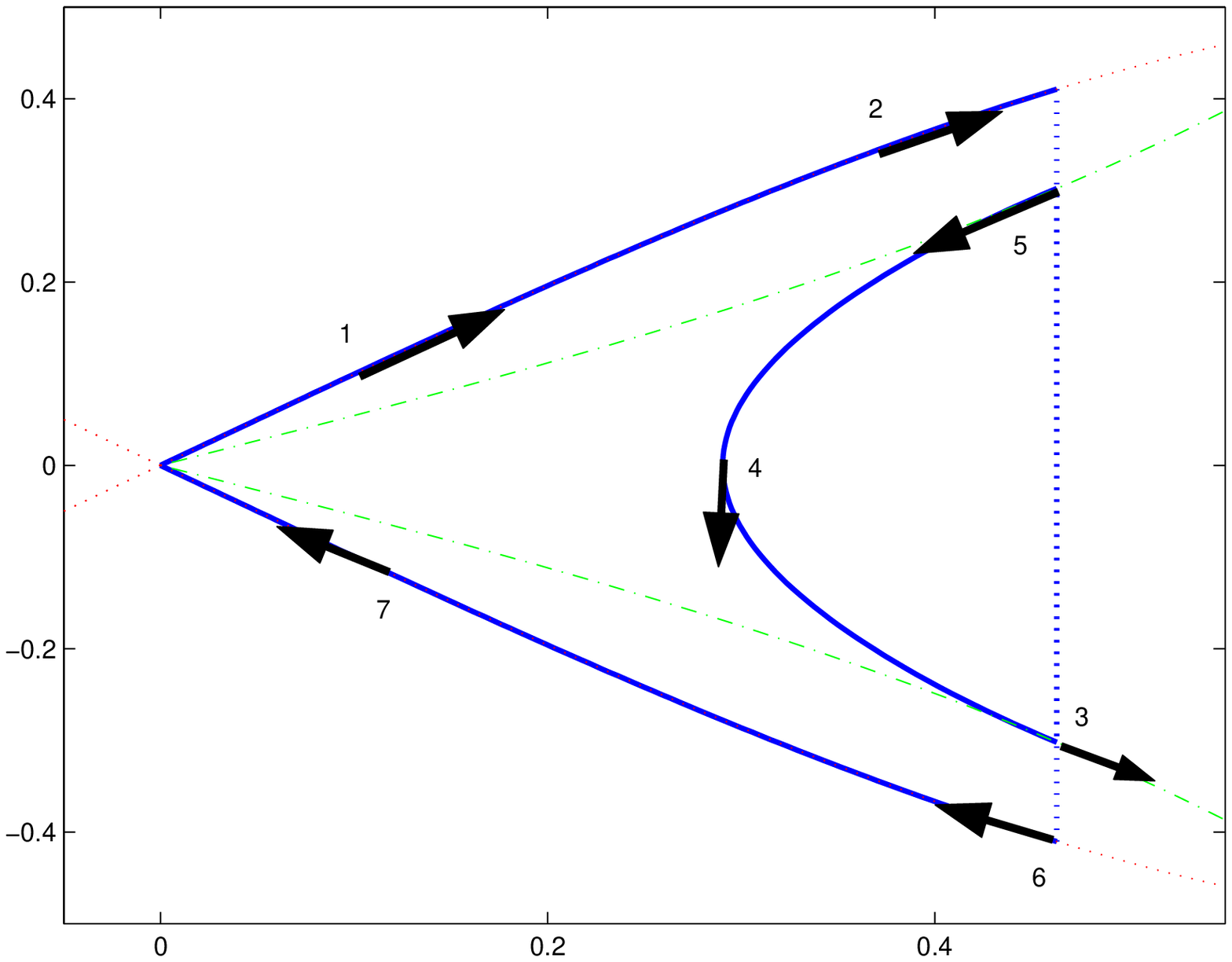} \\
\epsfxsize = 6.7cm \epsffile{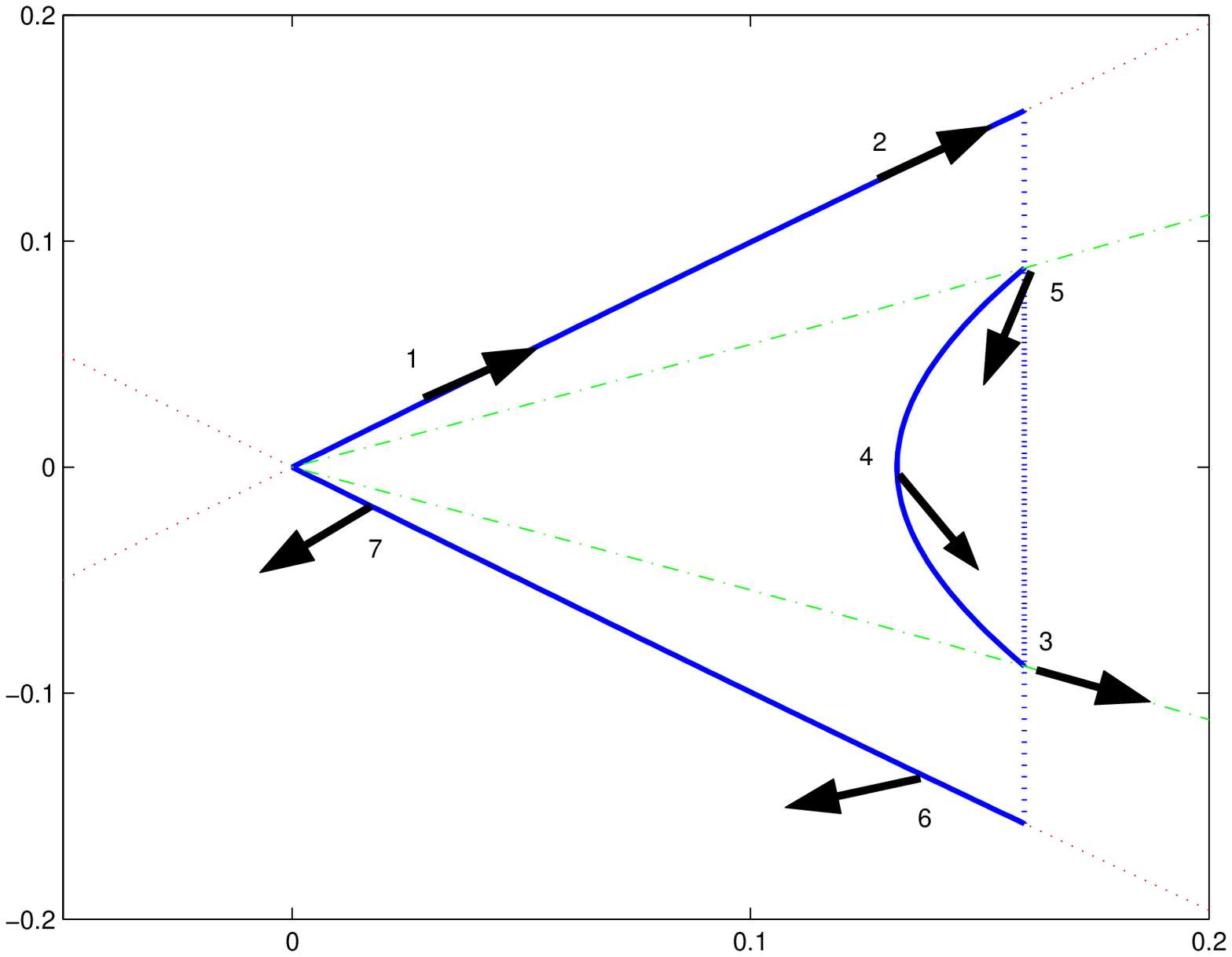} \\
\epsfxsize = 6.7cm \epsffile{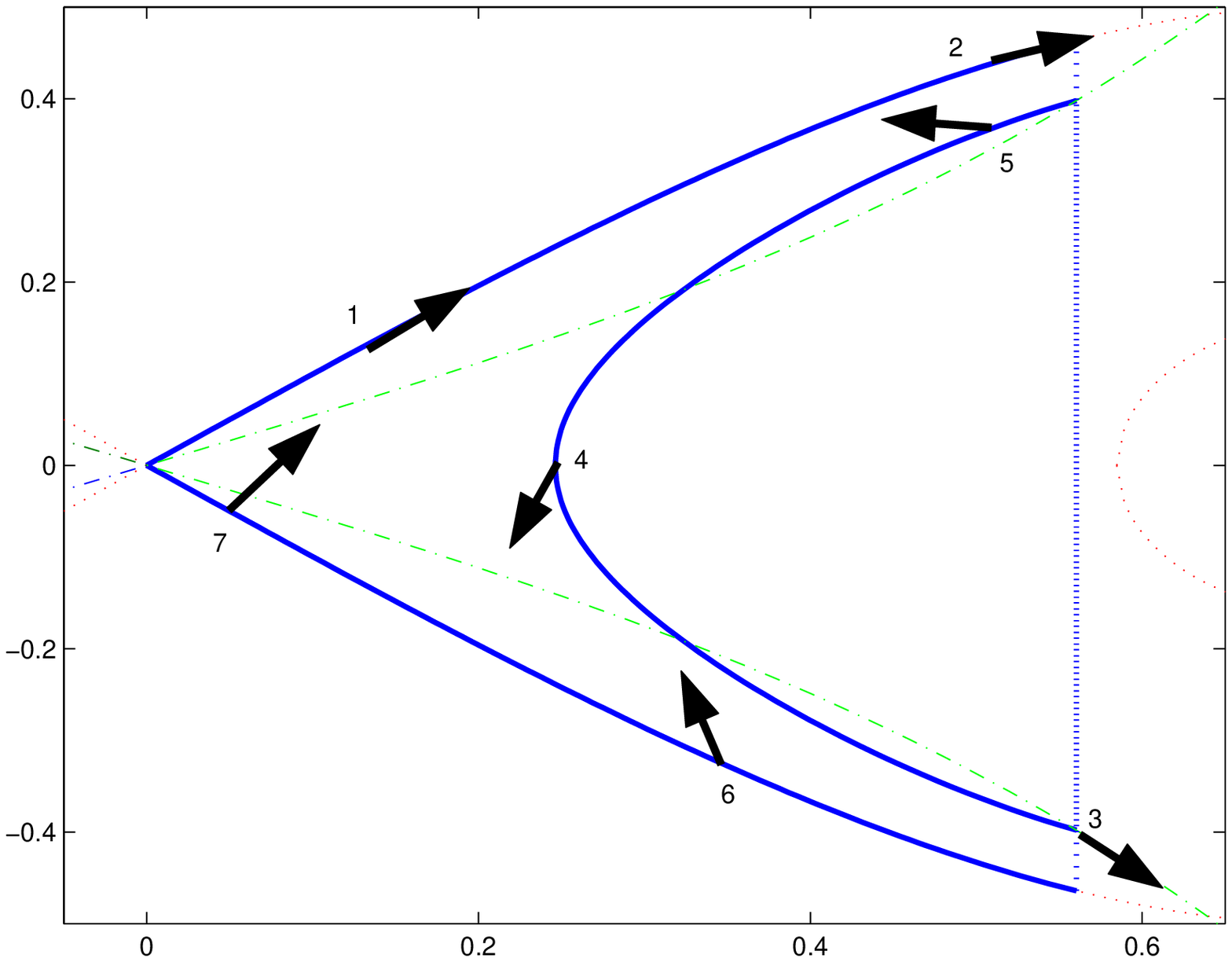}
\end{tabular}
%
\end{center}
\caption{The evolution of the tangent vector is shown for solutions with varying values of $L$.
In the first figure $L=L_{\rm thresh}$ and the evolved vector is parallel to the solution
throughout its evolution.  In the second figure $L<L_{\rm thresh}$ and the vector moves through
a smaller angle at each defect and we see $P \leq 1$.  
When $L>L_{\rm thresh}$ as in the third figure, the vector moves through a larger 
angle at each defect and forces $P \geq 2$.
 }
\label{fig:arrows}
\end{figure}

\noindent 
\begin{proof}
The result follows simply by evolving a vector 
(initialized as a tangent vector to the upper branch of the homoclinic)
around each standing wave under the influence of the equation of
variations.  This can be visualized by noting that, at the first defect, the
tangent vector to the unstable manifold $W^u$ is taken to a vector
that is tangent to the curve $J(W^u)$.  The observation of instability is 
then reduced to simply noting the manner in which this landing curve $J(W^u)$ 
intersects the periodic curve in the phase plane representing
the standing wave between the defects.
\begin{enumerate}

\item[1.]
At the threshold value $L = L_{\rm thresh}$ itself, the curve
$J(W^u)$ is tangent to the transient to which the solution jumps 
at the first well.  The variational solution, then, will still 
be tangent to the transient curve (although pointing backwards).
Hence, we can continue to trace the solution \emph{exactly} by 
following the tangent vector around this transient to the second jump.  
Then, as symmetry dictates, this vector is taken to a tangent vector 
to $W^s$ at the second jump.  This vector is seen to pass through
verticality only once (at $t=0$) and hence has $P = 1$. 

\item[2.]
For $L < L_{\rm thresh}$, the vector jumps through a smaller angle
at each well than it does at the bifurcation point itself.  At the first defect,
it then points along the curve $J(W^u)$, which in this case points \emph{up and into} 
the internal ellipse.  Because solutions are unique, the tangent vector 
to this ellipse provides a bound for our solution as it evolves, and the vector 
in question must retain its orientation with respect to the tangent to the ellipse; 
in particular, it rotates clockwise and must be pointing ``below''
the ellipse at the second defect.  However, the vector that is taken
to the curve $W^s$ is the vector tangent to $J^{-1}(W^s)$, and comparing
these two vectors we quickly see that $P \leq 1$ in this case.

\item[3.]
For $L < L_{\rm thresh}$, the vector jumps through a larger angle
at the first defect and now points along $J(W^u)$ \emph{down and out of} 
the internal ellipse, and so the tangents to the ellipse now provide a 
bound on the other side.  At the second well, the solution must still point
out of the ellipse and comparison with the vector tangent to $J^{-1}(W^s)$
forces $P \geq 2$ in this case.
\end{enumerate}

\noindent These three cases are sketched in figure \ref{fig:arrows}.
\end{proof}

\appendix
\section{Wellposedness of Gross-Pitaevskii / NLS}
\label{sec:wellposedness}

\noindent {\it Structural properties of NLS:}
\ The Gross-Pitaevskii / NLS equation is a  Hamiltonian
system, which can be written in the form:
\be
\mathrm{i} u_t\ =\ \frac{\delta}{\delta \bar u} \cH_{\mathrm{GP}}[u,\bar u],\ \label{eq:NLS-hamform}
\ee
where $\cH_{\mathrm{GP}}[u,\bar u]$ denotes the Hamiltonian (\ref{eq:GPenergy}).
Invariance with respect to time-translations implies that $\cH_{\mathrm{GP}}[u,\bar u]$
is conserved by the flow generated by \eqref{eq:NLS-hamform}. Additionally,
invariance under the transformation $u\mapsto e^{i\xi}u,\ \xi\in R^1$ implies
that $\cN[u,\bar u]$, the $L^2$ norm defined in (\ref{eq:Ndef}), is conserved.

For the spatially translation invariant case,  NLS has the
Galilean invariance:
\be
u(x,t)\ \mapsto\ u(x-st,t)\ \mathrm{e}^{\mathrm{i}(xv -\frac{1}{2}s^2t)}, \ s\in R^1.
\label{eq:Galilei}
\ee

\noindent{\it Well-posedness theory:}\ 
We present a sketch of the existence and uniqueness of solutions to (\ref{eq:GP});
 see also \cite{GHW:03}.  
The functionals $\cH_{\mathrm{GP}}[\cdot]$ and $\cN[\cdot]$ are well defined on
$H^1(R^1)$, the space of functions $f$, for which $f$ and $\D_xf$ are
square integrable.  It is therefore natural to construct the flow for
initial data of class $H^1$.  In fact, it can be shown that, for
initial conditions $u_0=u(x,t=0)\in H^1(R^1)$, there exists a unique
global solution of NLS, $u\in C^0(R^1;H^1(R))$,  in the sense of the
equivalent integral equation:
\begin{subequations}
\begin{gather}
u(t)\ =\  U(t)u_0\ +2\mathrm{i} \int_0^t U(t-s)\  
 |u(s)|^2u(s)\ \mathrm{d}s, \label{eq:NLS-ie}\\
U(t)\ \equiv\ 
 \exp\left(-\mathrm{i}Ht\right),\ H\equiv -\D_x^2+v(x). 
 \label{eq:propagator}
\end{gather}
\end{subequations}
The spectral decomposition of $H$ is known explicitly \cite{AGHH:88} and can
be used to construct $U(t)$ explicitly.

To show the existence of a solution to \eqref{eq:NLS-ie} in $H^1$, we must
show the existence of a $C^0(R^1;H^1(R))$ fixed point of the mapping
$u(x,t)\mapsto J[u](x,t)$, given by the right hand side of \eqref{eq:NLS-ie}.

We now outline the key ingredients of the proof.  To bound $J[u]$ and its
first derivative in $L^2$, we introduce the operator $\cA=I+P_cH$, where $P_c$
denotes the projection onto the continuous spectral part of $H$. Note that
$\cA$ is a nonnegative operator, since the continuous spectrum of $H$ is the
nonnegative real half-line.  Moreover, we expect $\|\cA^\frac{1}{2}f\|\ \sim
\|f\|_{H^1} \equiv\ \|(I-\D_x^2)^\frac{1}{2}f\|_{L^2}$. In fact, we shall use
that the following operators are bounded from $L^2$ to $L^2$: 
\be
\cA^\frac{1}{2}(I-\D_x^2)^{-\frac{1}{2}},\
\cA^{-\frac{1}{2}}(I-\D_x^2)^\frac{1}{2} .  \nonumber 
\ee 
This follows from the boundedness of the {\it wave operators} on 
$H^1$ \cite{Weder:99}.  Therefore, we have an equivalence of norms 
\be C_1\ \| f\|_{H^1}\ \le\ \|
\cA^\frac{1}{2}f\|_{L^2}\ \le\ C_2\ \| f\|_{H^1} .
\label{eq:norm-equiv}
\ee
Our formulation \eqref{eq:NLS-ie} and introduction of $\cA$ are related
to the nice property that $\cA$, and hence also functions of $\cA$,
commute with the propagator $\exp(-\mathrm{i}Ht)$. We shall also use the
Sobolev inequality:
\be 
|f(x)|^2 \le\  C\|f\|_{L^2}\|\D_xf\|_{L^2} ,
\label{eq:sobolev}
\ee
and the Leibniz rule\ \cite{KP:88}:
\be
\|(I-\D_x^2)^\frac{1}{2}(fg)\|\ \le 
\ C\left( \| f\|_{L^\infty}\ \|(I-\D_x^2)^\frac{1}{2}g\|_{L^2}\ +
 \ \|(I-\D_x^2)^\frac{1}{2}f\|_{L^2}\ \| g\|_{L^\infty} \right) .
\label{eq:leibniz}
\ee

\noindent
Since $U(t)$ is unitary in $L^2$, we have
\begin{align}
&\| \cA^\frac{1}{2} J[u]\|_{L^2}\ \nn\\
&\le\ \|\cA^\frac{1}{2} u_0\|_{L^2}\ +\ 
 \g\int_0^t \|  \cA^\frac{1}{2} |u(s)|^2u(s) \|_{L^2}\ ds\nn\\ 
&=\ \|\cA^\frac{1}{2} u_0\|_{L^2}\ +\ 
 \g\int_0^t \|  \left(\cA^\frac{1}{2}(I-\D_x^2)^{-\frac{1}{2}}\right)\cdot
 \left( (I-\D_x^2)^\frac{1}{2} |u(s)|^2u(s)\right) \|_{L^2}\ ds.
\label{eq:run1}
\end{align}
By \eqref{eq:norm-equiv}, \eqref{eq:sobolev}, and \eqref{eq:leibniz},
\be
\|J[u](t)\|_{H^1}\ \le\   
  C\| \cA J[u]\|_{L^2}\ \le\ C_1\|u_0\|_{H^1}\ +\ 
 C_2\ T\  
 \sup_{s\in [0,T]} \|u(s)\|_{H^1}^3 .
\label{eq:Jbound}
\ee
Now assume that $u$ is such that $\sup_{s\in [0,T]}\|u(s)\|_{H^1} \le 2C_1$.
Then, by \eqref{eq:Jbound},  choosing $T<T_1$ sufficiently small,
 $\sup_{s\in [0,T]}\|J[u](s)\|_{H^1} \le 2C_1\|u_0\|_{H^1}$ and so
\be
\|J[u](t)\|_{H^1}\ \le\ C_1\|u_0\|_{H^1}\ +\ C_2\ T\ \left( 2C_1\|u_0\|_{H^1}
\right)^3 .
\nonumber\ee
It follows that for $0<T<T_1$ sufficiently small, the transformation
$J[\cdot]$ maps a ball $C^0([0,T];H^1(R))$ into itself. 
A similar calculation shows that 
\be
\|J[u](t)\ -\ J[v](t)\|_{H^1}\ \le\  
 K\ T\ \left(2C_1\|u_0\|_{H^1}\right)^2\ 
 \sup_{s\in [0,T]}\ \| u(s)-v(s)\|_{H^1} ,  
\ee
and therefore for $0<T<T_2\le T_1$, the transformation $J[\cdot]$
is a contraction on this ball. Therefore,  $J[\cdot]$ has a unique fixed point
in $C^0([0,T];H^1(R))$ for $T$ sufficiently small and local existence in time
of the flow follows.  Global existence in time follows from the {\it \'a
priori} bound on the $H^1$ norm of the solution implied by the
time-invariance of $L^2$ norm, $\cN$, and the Hamiltonian, $\cH_{\mathrm{GP}}$. 
\medskip

\nit{\bf Acknowledgements}

\nit This work was initiated while R.K. Jackson participated in the Bell Labs / Lucent
 Student Intern Program. R.K. Jackson would like to thank C.K.R.T. Jones for fruitful discussions 
 and guidance, and also gratefully acknowledges the current support of the National Science Foundation
 under award  DMS-0202542.


\end{document}